\documentclass{elsart}

\usepackage{amssymb,graphicx}

\newcommand{\URAD}{TaylUR}
\newcommand{\FOR}{Fortran~95}
\newcommand{\taylor}{{\tt taylor}}
\newcommand{\wrt}{w.r.t.}

\newcounter{bla}
\newenvironment{refnummer}{%
\list{[\arabic{bla}]}%
{\usecounter{bla}%
 \setlength{\itemindent}{0pt}%
 \setlength{\topsep}{0pt}%
 \setlength{\itemsep}{0pt}%
 \setlength{\labelsep}{2pt}%
 \setlength{\listparindent}{0pt}%
 \settowidth{\labelwidth}{[9]}%
 \setlength{\leftmargin}{\labelwidth}%
 \addtolength{\leftmargin}{\labelsep}%
 \setlength{\rightmargin}{0pt}}}
 {\endlist}

\begin{document}
\begin{frontmatter}


\title{\URAD, an arbitrary-order diagonal automatic differentiation
  package for \FOR}

\author{G.M. von Hippel\thanksref{a}}
\thanks[a]{Corresponding author}
\address{Department of Physics, University of Regina, Regina,
  Saskatchewan, S4S 0A2, Canada}

\ead{vonhippg@uregina.ca}
\ead[url]{http://uregina.ca/\~\/vonhippg/}


\begin{abstract}
We present \URAD\/, a \FOR\/ module to automatically compute the
numerical values of a complex-valued function's derivatives with
respect to several variables up to an arbitrary order in each
variable, but excluding mixed derivatives. Arithmetic operators and
Fortran intrinsics are overloaded to act correctly on objects of a
defined type {\tt taylor}, which encodes a function along with its
first few derivatives with respect to the user-defined independent
variables. Derivatives of products and composite functions are
computed using Leibniz's rule and Fa\`a di Bruno's formula. \URAD\/
makes heavy use of operator overloading and other \FOR\/ features such
as elemental functions.

\end{abstract}

\end{frontmatter}


{\bf PROGRAM SUMMARY}

\begin{small}
\noindent
{\em Manuscript Title:} TaylUR, an arbitrary-order diagonal automatic
differentiation package for Fortran 95                        \\
{\em Authors:} G.M. von Hippel                                \\
{\em Program Title:} TaylUR                                   \\
{\em Journal Reference:}                                      \\
{\em Catalogue identifier:}                                   \\
{\em Licensing provisions:} none                              \\
{\em Programming language:} Fortran 95                        \\
{\em Computer:}  Any computer with a conforming Fortran 95
                 compiler                                     \\
{\em Operating system:} Any system with a conforming
                        Fortran 95 compiler                   
{\em Classification:} 4.12 Other Numerical Methods,
                      4.14 Utility                            \\

{\em Nature of problem:}\\
  Problems that require potentially high orders of derivatives with
  respect to some variables, such as e.g. expansions of Feynman
  diagrams in particle masses in perturbative Quantum Field Theory,
  and which cannot be treated using existing Fortran modules for
  automatic differentiation [1-2]. \\
   \\
{\em Solution method:}\\
  Arithmetic operators and Fortran intrinsics are overloaded to act
  correctly on objects of a defined type {\tt taylor}, which encodes a
  function along with its first few derivatives with respect to the
  user-defined independent variables. Derivatives of products and
  composite functions are computed using Leibniz's rule and F\`aa di
  Bruno's formula. \\
   \\
{\em Restrictions:}\\
  Memory and CPU time constraints may restrict the number of variables
  and Taylor expansion order that can be achieved. Loss of numerical
  accuracy due to cancellation may become an issue at very high
  orders. \\
   \\
{\em Unusual features:}\\
  No mixed higher-order derivatives are computed. The complex
  conjugation operation assumes all independent variables to be real. \\
   \\
{\em Running time:}\\
  The running time of TaylUR operations depends linearly on the number
  of variables. Its dependence on the Taylor expansion order varies
  from linear (for linear operations) through quadratic (for
  multiplication) to exponential (for elementary function calls). \\
   \\
{\em References:}
\begin{refnummer}
\item 
  C.~W. Straka,
  ADF95: Tool for automatic differentiation of a FORTRAN code designed
  for large numbers of independent variables,
  Comput.~Phys.~Commun. {\bf 168} (2005) 123-139
  [arXiv:cs.MS/0503014].
\item
  S. Stamatiadis, R. Prosmiti, S.~C. Farantos,
  {\sc auto\_deriv}: Tool for automatic differentiation of a FORTRAN
  code,
  Comput.~Phys.~Commun. {\bf 127} (2000) 343-355.
\end{refnummer}

\end{small}

\newpage


\hspace{1pc}
{\bf LONG WRITE-UP}

\section{Introduction}
\label{sect:intro}

There has recently been an increased interest in the physics
literature in methods that allow to automatically compute the
numerical values of the derivatives of a function along with the
function itself \cite{hart,morningstar,straka}. Since the need to do this
arises in different contexts, different methods have been proposed and
implemented. In many cases, such as in differential equation solvers,
only the first and possibly second derivatives are needed, a task for
which a number of tools exist \cite{straka,stamatiadis,autodiff}.
There are, however, some cases, such as the expansion of Feynman
diagrams in external momenta or particle masses and other series
expansions in Quantum Field Theory, where at least some of the
higher-order derivatives up to some relatively large order are needed,
and few tools to deal with this situation were available so
far. It is thus the purpose of this paper to present \URAD, a \FOR\/
module that addresses this need by providing a mechanism to compute
the values of the higher derivatives of a function along with the
function itself.

In the field of automatic differentiation \cite{autodiff}, two
different approaches are generally distinguished: Source
transformation methods, which take an existing code for the
computation of a function as their input and produce a code that
computes its derivatives as output, and operator overloading methods,
which make use of operator overloading and other object-oriented
features of a language to encapsulate the task of computing
derivatives within an object that has the same ``user interface'' as a
real or complex number. While source transformation has the advantage
of producing faster code, since it can take full advantage of a
compiler's optimising features, operator overloading provides for an
easier and more convenient user interface, in particular since no
additional step is inserted into the compile-link-run cycle. \URAD\/
uses an operator overloading strategy to provide a new type \taylor\/
that acts like a complex (or real) number while containing the values
of the derivatives of the function it represents along with that
function's value.

In order to emulate the behaviour of intrinsic numerical data types as
closely as possible, \URAD\/ makes significant use of the \FOR\/
features for pure and elemental functions. Thus very few, if any,
changes to existing user code (apart from declaring objects to be of
type \taylor) will be necessary.

The intended area of application of \URAD\/ is fairly orthogonal to
that of other, existing systems like ADF95 \cite{straka} or
{\sc auto\_deriv} \cite{stamatiadis}: Where the latter are aimed
primarily at use in implicit differential equation solvers, where only
first and possibly second order derivatives are needed, but efficient
handling of large numbers of variables with potentially sparse
Jacobian and Hessian matrices is a primary goal, \URAD\/ aims at cases
such as Feynman diagram differentiation, where the number of variables
is usually limited, but an expansion to higher orders is
needed. \URAD\/ performs well with large numbers of variables, but
does not exploit any existing sparsity structure of the derivatives.

There are some areas of application, such as optimization problems or
higher-dimensional interpolation, which require mixed higher-order
derivatives, something which neither \URAD\/ nor ADF95 or {\sc
  auto\_deriv} provide.

Where it is acceptable to link a \FOR\/ program against code
written in C++, the C++ package ADOL-C \cite{griewank,griewank2} is
available. ADOL-C uses a very different approach internally than
\URAD\/, employing a sophisticated storage allocation system to
maintain a data structure that allows the computation of higher
derivative tensors, including mixed derivatives, for functions written
in C/C++. The advantage of computing mixed higher-order derivatives
which ADOL-C has over \URAD\/ is bought at the expense of numerous
memory accesses, which significantly slow down the code at runtime,
although on the other hand it allows for better scaling behaviour in
the Taylor expansion order. The beam dynamics simulation and analysis
code COSY INFINITY \cite{makino} and the general-purpose numerical
Harwell Subroutine Library \cite{hsl} also contain arbitrary-order
automatic differentiation code that computes mixed higher-order
derivatives and can be linked against \FOR\/ programs.

We believe that the straightforward interface and relatively small
code size (about a quarter the size of ADOL-C) of \URAD\/, as well as
the fact that it is coded entirely in \FOR, offer a fair balance for
the lack of mixed higher-order derivatives, and make it very
competitive with these larger libraries for those applications where
mixed higher-order derivatives are not required.

It should also be noted that the existing packages for
automatic differentiation using operator overloading have so far been
restricted to use real-valued functions only. \URAD\/ overcomes this
limitation and provides a complex-valued function type, which is
commonly needed e.g. in the evaluation of Feynman diagrams in
high-energy physics.

\section{Detailed description of program}
\label{sect:descr}

\subsection{Usage of the program}
\label{subs:usage}

In order to employ the \URAD\/ module, the user must include a {\tt
USE} statement for \URAD\/ at the beginning of his program. In
addition to this, any variable for which the computation of
derivatives is desired, as well as all variables and user functions
which feed into its computation over the course of the user program,
must be declared to be of {\tt TYPE(taylor)}.

Independent variables are created by the function {\tt INDEPENDENT},
which takes the index of the independent variable and its value as
arguments, as in the following example:
\begin{verbatim}
   TYPE(taylor) :: x,y,f

   x = independent(1,0.3)
   y = independent(2,2)
   f = y**2 -3*x
\end{verbatim}
which declares {\tt x} and {\tt y} to be the independent variables with
index 1 and 2, respectively, and assigns them the values {\tt 0.3} and
{\tt 2}, before computing the function {\tt f} from them. The function
value of {\tt f} will be {\tt 3.1}, its first derivative with respect
to (\wrt\/) the
first independent variable will be {\tt -3.} with all higher
derivatives \wrt\/ this variable vanishing, and its first derivative
\wrt\/ the second independent variable will be {\tt 4.}, the second
derivative {\tt 2.}, all higher derivatives \wrt\/ this variable vanish
again, as do any derivatives \wrt\/ other variables.

As a slightly more realistic example let us look at the computation of
the wavefunction renormalisation constant in Euclidean scalar $\phi^3$
theory. At the one-loop level, we need to compute
$\frac{\partial^2}{\partial p_4^2}\Pi(p)|_{p_4=im}$, where $\Pi(p)$ is
given by the bubble diagram. A way to code this might be given by
\begin{verbatim}
  FUNCTION bubble_diagram(k,p_,m_)

  USE TaylUR

  COMPLEX :: p_(4),bubble_diagram
  REAL :: k(4),m_

  TYPE(taylor) :: p(4),m,feynman
  INTEGER :: mu

  DO mu=1,4
    p(mu) = independent(mu,p_(mu))
  ENDDO
  m = independent(5,m_)

  k(4) = k(4) - 0.5*p(4)  ! Shift k_4 countour
  feynman = 1/(sum(k**2)+m**2)/(sum((k+p)**2)+m**2)

  bubble_diagram = derivative(feynman,4,2)/(2*Pi)**4

  END FUNCTION bubble_diagram
\end{verbatim}
This routine might then be passed to an integration routine which
expects a function with the given interface and performs an
appropriately regularised integration.

While in this case the derivative can be easily computed analytically,
this is no longer the case in e.g. lattice perturbation theory, where
(especially for improved actions \cite{hart}) the Feynman rules quickly
become too complex to allow analytical calculations to be performed
easily. It should also be noted that changing just the two last
arguments of the {\tt DERIVATIVE} function call will allow to compute
the mass dependence of the mass renormalisation. With an
appropriately implemented integration method, it is even possible to
numerically integrate a \taylor\/ object-valued function for a \taylor\/
object-valued result.

\subsection{Structure and handling of \taylor\/ objects}
\label{subs:type}

The \taylor\/ type is defined internally as
\begin{verbatim}
TYPE taylor
 COMPLEX(kind=dc_kind) :: drv(1:N_taylor_vars,0:Max_taylor_order)
END TYPE taylor
\end{verbatim}
where {\tt dc\_kind} is defined as the kind parameter of a double
precision {\tt complex}. The field {\tt drv(i,n)} holds the {\tt n}-th
derivative \wrt\/ the {\tt i}-th independent variable, where any
zeroth derivative is defined as equal to the function value.

The maximal order of the Taylor expansion is determined by the module
parameter {\tt Max\_taylor\_order}, the total number of variables by
the module parameter {\tt N\_taylor\_vars}. Where needed, these
parameters can be changed to provide for higher orders or more
variables, or to speed up code by reducing the
numbers.\footnote{Memory constraints may impose an upper limit of both
{\tt Max\_taylor\_order} and {\tt N\_taylor\_vars}, as may compiler
restrictions (at least some versions of the Intel Fortran compiler
will cause a segmentation fault if {\tt Max\_taylor\_order} exceeds
about 12, and some Sun compilers experience arithmetic exceptions at
about the same order).}

Where a specific portion of a program does not need the full Taylor
expansion up to order {\tt Max\_taylor\_order}, it is possible to set
the module variable {\tt Taylor\_order} to a lower value in order to
compute only derivatives up to that order.

Independent variables are created by the function {\tt INDEPENDENT},
which takes the index of the independent variable and its value as
arguments. It is the user's responsibility to make sure that no two
\taylor\/ objects are declared to share the same independent variable
index, since it is not possible for the \URAD\/ module to keep track
of this.

There are a number of user-defined functions for accessing the
value and derivatives of the function encoded by a \taylor\/ object.
{\tt VALUE(t)} returns the value of the \taylor\/ object {\tt t}, whereas
{\tt DERIVATIVE(t,i,n)} gives the {\tt n}-th partial derivative of the
\taylor\/ object {\tt t} \wrt\/ the {\tt i}-th independent variable. The
expansion $(f,\partial_{x_i}f,\partial^2_{x_i}f,\dots)$ \wrt\/ can be
obtained as an array by using {\tt EXPANSION(t,i)}.

It is also possible to obtain the vector of first partial derivatives
as an array by using the function {\tt GRADIENT(t)}, and the Laplacian
as {\tt LAPLACIAN(t)}.

\subsection{Overloaded operations on \taylor\/ objects}
\label{subs:overlop}

The assignment operator {\tt =} has been overloaded to allow
assignment of intrinsic types to \taylor\/ objects and vice versa. It
should be noted that, in accordance with the standard behaviour of
\FOR\/ intrinsic types, assignment to a REAL will result in the real
part of the right-hand side being taken implictly.

All standard \FOR\/ arithmetic operators ({\tt +},{\tt -},{\tt *},
{\tt /},{\tt **}) have been overloaded to act on \taylor\/
objects. Numbers of both double and default precision real and complex
type as well as integers can be combined with \taylor\/ objects by addition,
subtraction, multiplication and division in any order, and \taylor\/
objects can be raised to integer powers. All these operations are
defined with the {\tt elemental} attribute and therefore can be used
with arrays of \taylor\/ objects using the usual \FOR\/ array syntax.

All comparison operators have been overloaded to allow comparison of
\taylor\/ objects with both double and default precision reals and
integers as well as with complex numbers, where this makes sense. The
comparison operators compare the value of the \taylor\/ objects only,
neglecting their derivatives. In addition to these intrinsic
operations, two user-defined comparison operators {\tt .IDENT.} and
{\tt .NIDENT.} exist, which check for identity and non-identity of
the complete \taylor\/ series, as opposed to the comparison of values
only carried out by the intrinsic operators {\tt ==} and {\tt /=}.
All comparison operators are elemental.

\subsection{Overloaded intrinsic functions of \taylor\/ objects}
\label{subs:overlin}

All \FOR\/ intrinsics which make sense on a Taylor-expanded quantity
and which can be fully implemented as user-defined functions, have
been overloaded to work correctly on \taylor\/ objects. Specifically,
the functions {\tt ABS}, {\tt ACOS}, {\tt AIMAG}, {\tt ASIN},
{\tt ATAN}, {\tt ATAN2}, {\tt CONJG}, {\tt COS}, {\tt COSH},
{\tt DIM}, {\tt EXP}, {\tt LOG}, {\tt LOG10}, {\tt MOD},
{\tt MODULO}, {\tt SIGN}, {\tt SIN}, {\tt SINH}, {\tt SQRT},
{\tt TAN}, {\tt TANH} accept \taylor\/ objects as their arguments, and
{\tt MATMUL} and {\tt DOT\_PRODUCT} accept arrays of \taylor\/ objects
as their arguments.

In the case of the functions {\tt REAL} and {\tt AIMAG}, a conscious
decision was made to have them behave differently from their intrinsic
counterparts in that they do \emph{not} convert to {\tt real} type,
but return an result of type \taylor\/ (with the real/imaginary part
of each derivative taken) instead. This was done so that the
mathematical functions $\Re$ (real part) and $\Im$ (imaginary part)
are available on \taylor\/ objects, and since assignment of a
\taylor\/ object to a {\tt real} variable will convert it to a {\tt real}
anyway, no functionality is lost. In the case where it is necessary to
assign the value of one \taylor\/ object to another \taylor\/ object
as a constant, or where the value of a \taylor\/ object has to be
passed to an external function that accepts only a {\tt complex} or
{\tt real} argument, the user-defined {\tt VALUE} function may be used
instead. A pair of user-defined functions {\tt REALVALUE} and
{\tt IMAGVALUE}, which return the real and imaginary part of the value
of a \taylor\/ object, respectively, are also provided. 

The following \FOR\/ intrinsics cannot be fully emulated by
user-defined functions, since they return results of different kinds
depending on the value of an argument, which is impossible to
achieve with a function written in \FOR\/: {\tt AINT}, {\tt ANINT},
{\tt CEILING}, {\tt FLOOR}, {\tt INT}, {\tt NINT}, {\tt REAL}. These
functions accept \taylor\/ objects as their arguments only when the
optional {\tt kind} argument is absent.

The \FOR\/ intrinsics {\tt MAX} and {\tt MIN} that accept arbitrary
variable numbers of arguments (which a function written in \FOR\/
cannot emulate) accept \taylor\/ objects as their arguments in their
two-argument form only.

The \FOR\/ array reduction intrinsics {\tt MAXLOC}, {\tt MAXVAL},
{\tt MINLOC}, {\tt MINVAL}, {\tt PRODUCT} and {\tt SUM} accept
arguments of variable rank along with an optional argument {\tt dim}
to denote the dimension along which reduction is to be performed. This
behaviour, too, cannot be emulated by a \FOR\/ function; these
functions accept \taylor\/ object arrays of rank one only. Otherwise
they act as their intrinsic counterparts, including the existence of
the optional {\tt mask} argument.

The \taylor\/ functions are elemental where their intrinsic
counterparts are.\footnote{As a reminder, an elemental function is one
which can be called with an array instead of a scalar passed as its
argument, and will return an array contain the result of its
application to each element of the passed array in turn.}

Those functions whose intrinsic counterparts are restricted to real or
integer arguments ({\tt ACOS}, {\tt AINT}, {\tt ANINT}, {\tt ASIN},
{\tt ATAN}, {\tt ATAN2}, {\tt CEILING}, {\tt DIM},
{\tt FLOOR}, {\tt INT}, {\tt LOG10}, {\tt MAX}, {\tt MAXLOC},
{\tt MAXVAL}, {\tt MIN}, {\tt MINLOC}, {\tt MINVAL}, {\tt MOD},
{\tt MODULO}, {\tt NINT}, {\tt SIGN}) will take the real part of a
\taylor\/ object first and should be applied to real-valued \taylor\/
objects only, just like their intrinsic counterparts would be applied
to {\tt real} numbers only. Depending on the value of the variable
{\tt Real\_args\_warn}, which defaults to {\tt .TRUE.}, these functions
will warn about a complex argument (with imaginary part greater than
{\tt Real\_args\_tol}) being passed by returning a {\tt NaN} (not a
number) value, or $\pm${\tt HUGE} in the case of integer functions;
this behaviour can be turned off where desired by setting
{\tt Real\_args\_warn} to {\tt .FALSE.} in the user's code.

On the other hand, {\tt COSH}, {\tt SINH}, {\tt TAN} and {\tt TANH}
work correctly with complex-valued \taylor\/ objects, although their
intrinsic counterparts are (somewhat arbitrarily) restricted to
{\tt real} arguments.

It should be noted that, while \URAD\/ accepts complex-valued
independent variables, the {\tt CONJG} function assumes that all
independent variables are real. In particular, no attempt is made to
implement any features of Wirtinger calculus.

In those cases where the derivative of a function becomes undefined at
certain points (as for {\tt ABS}, {\tt AINT}, {\tt ANINT}, {\tt MAX},
{\tt MIN}, {\tt MOD}, {\tt MODULO} and {\tt SQRT}), while the value is
well defined, the derivative fields will be filled with {\tt NaN} (not
a number) values by assigning them to be {\tt 0./0.} Depending on the
compiler and system settings, this may cause the program to stop.

Examples of the usage of all routines can be found in the test program
distributed with \URAD.

\subsection{Mathematical details of implementation}
\label{subs:math}

The derivatives of products of \taylor\/ objects are computed using
Leibniz's rule
\begin{equation}
\frac{\partial^n}{\partial x_i^n}(fg) = 
\sum_{k=0}^n {n \choose k} \left(\frac{\partial^k f}{\partial
  x_i^k}\right)\left(\frac{\partial^{n-k} g}{\partial x_i^{n-k}}\right)
\end{equation}
Leibniz's rule is also employed to compute derivatives of quotients
and square roots by using the equalities
\begin{equation}
0 = \frac{\partial^n}{\partial x_i^n}(1) = 
\frac{\partial^n}{\partial x_i^n}(ff^{-1}) =
\sum_{k=0}^n {n \choose k} \left(\frac{\partial^k f}{\partial
  x_i^k}\right)\left(\frac{\partial^{n-k} f^{-1}}{\partial x_i^{n-k}}\right)
\end{equation}
\begin{equation}
\frac{\partial^n}{\partial x_i^n}f = \frac{\partial^n}{\partial
  x_i^n}(\sqrt{f})^2 = \sum_{k=0}^n {n \choose k}
  \left(\frac{\partial^k \sqrt{f}}{\partial x_i^k}\right)
\left(\frac{\partial^{n-k} \sqrt{f}}{\partial x_i^{n-k}}\right)
\end{equation}
and solving for the $n$-the derivative of $f^{-1}$ or $\sqrt{f}$,
respectively. 

\URAD\/ differs significantly from similar tools such as
{\sc auto\_deriv} \cite{stamatiadis} or ADF95 \cite{straka} that use
hard-coded chain rule expressions for the derivatives of
intrinsics. Such an approach is obviously unsuitable for the
arbitrary-order case. \URAD\/ instead uses Fa\`a di Bruno's formula
\cite{faadibruno,faadibruno2} for the $n$-th derivative of a composite
function:
\begin{equation}
\frac{\partial^n}{\partial x_i^n} F(y(x)) =
\sum_{\{k\}} \frac{n!}{\prod_\nu k_\nu!\nu!^{k_\nu}}
\frac{\d^{|k|} F}{\d y^{|k|}}
\prod_{\mu=1}^{n} \left(\frac{\partial^\mu y}{\partial x_i^\mu} \right)^{k_\mu}
\end{equation}
where $|k|=\sum_\mu k_\mu$, and the sum runs over all integer vectors
$k$ that satisfy the conditions $0\leq k_\mu \leq n$ and $\sum_\mu \mu
k_\mu = n$.

A subroutine {\tt FDB\_GENERATE} that generates and stores the needed
vectors $k$ along with the precomputed weight of each term
$F^{(m)}\prod_\mu (y^{(\mu)})^{k_\mu}$ is called from within the
{\tt INDEPENDENT} function. Those functions requiring the
computation of derivatives via Fa\`a di Bruno's formula call another
function {\tt FDB} which returns these precomputed values. This
separation allows to compute the vectors only once and store them for
better performance, while maintaining the {\tt elemental} status of
the overloaded intrinsics (which requires them to be {\tt pure},
prohibiting any side-effects such as manipulating external data or
{\tt save} variables).

\subsection{Typical running time and memory usage}
\label{subs:run}

The runtime memory and CPU requirements of the \URAD\/ package are
determined by the {\tt N\_\taylor\_vars}, {\tt Max\_taylor\_order}
parameters and, to some extent, the {\tt Taylor\_order} variable.

A \taylor\/ object requires {\tt N\_taylor\_vars}$\times$
{\tt Max\_taylor\_order} times the memory that a complex variable of
kind {\tt dc\_kind} requires, and this size is fixed at compilation
time.

The time taken to perform operations on \taylor\/ objects depends on
the operation, the compiled value of {\tt N\_taylor\_vars} and the
run-time value of {\tt Taylor\_order}. All operations scale
approximately linearly in the number of variables, although this
partially depends on the CPU and compiler, since the dependence on
{\tt N\_taylor\_vars} is through array assignments alone. The run-time
of linear operations, such as assignment, addition, multiplication
with a constant or comparison for identity, as well as some of the
simpler mathematical functions like {\tt MOD}, will scale
approximately linearly with {\tt Max\_taylor\_order} (again through
array assignment), while being independent of
{\tt Taylor\_order}. Multiplication and division of \taylor\/ objects,
as well as the {\tt SQRT} function, will have a run-time scaling
quadratically with {\tt Taylor\_order}, while the run-time of the
{\tt EXP} function, logarithms, trigonometric, hyperbolic and inverse
trigonometric functions will scale approximately exponentially with
{\tt Taylor\_order}, due to the increase in the number of terms
combining different lower-order derivatives of their arguments that
need to be combined.

\subsection{Limitations on the complexity of the problem}
\label{subs:limits}

Besides the limits on the number of variables and maximal expansion
order that memory and CPU time constraints may impose, there are
limits on the expansion order that may be achieved due to the finite
precision of floating-point operations taken in conjunction with the
large number of potentially large terms that need to be added to
obtain the value of high-order derivatives.

This becomes a problem in particular when functions whose $n$-th
derivatives are large are combined to form a function with a small
$n$-th derivative, or when contributions from different orders combine
in a product of quotient to yield a small result for some higher-order
derivative of the result. In these cases, the resulting value for the
derivative can be many orders of magnitude less accurate than the
value obtained for the function value or lower-order derivatives.

In order to avoid a potential total loss of precision, users
interested in high-order derivatives should monitor the derivatives of
intermediate results and compare them to the derivatives of the final
answer. If the derivatives of the final result become insignificant
when compared to those of the intermediate steps that entered its
computation, the final result is likely to be dominated by numerical
noise.

\subsection{Unusual features of the program}
\label{subs:unfeat}

Apart from any language-imposed limitations in the emulation of the
behaviour of certain \FOR\/ intrinsics mentioned in section
\ref{subs:usage}, \URAD\/ is limited in that it does not attempt to
compute any mixed second or higher-order derivatives. This limitation
is imposed for a number of reasons: Firstly, the memory and time needed
for the storage and computation of mixed higher-order derivatives gets
out of hand very quickly as one goes up to higher orders. Secondly,
the computation of mixed derivatives of arbitrary order involves a
programming effort that is rather disproportionate to the use that is
likely to be made of them in most applications.

We would also like to repeat that \URAD\/ allows for complex-valued
functions and variables, while those functions that are only defined
for real arguments (in particular those that refer to ordering
relations) will silently take the real parts of their arguments. It
also bears repeating that the {\tt CONJG} function assumes that all
independent variables are real-valued, and that no attempt is made to
implement Wirtinger calculus.

\subsection{Testing and verification}
\label{subs:tests}

The \URAD\/ package has been tested on a range of computer systems
(Linux/ Intel, Linux/Alpha and SunOS/Sparc) with a number of different
compilers (Intel, Compaq and Sun).

The test suite used to test \URAD\/ for potential bugs and errors
consists of both testing the derivatives computed for a number of
functions with known derivative expansions against their analytically
determined values and of various sanity checks such as that {\tt f/f}
is equal to one with all derivatives vanishing to within a reasonable
accuracy. 

\URAD\/ has evolved out of similar, more restricted codes used by
the author in the automatic differentiation of Feynman diagrams in
perturbative lattice QCD \cite{hart} and chiral perturbation theory on
the lattice. These codes had been tested and used in practice and have
been found to work correctly, giving results in agreement with
analytical results whenever those were available. While \URAD\/
contains some significant extensions compared to these codes, in
particular the intrinsic functions whose derivative expansion is based
of Fa\`a di Bruno's formula, these extensions have been well tested
and found to be stable.

\begin{figure}
\includegraphics[height=0.49\textwidth,keepaspectratio=,angle=270]{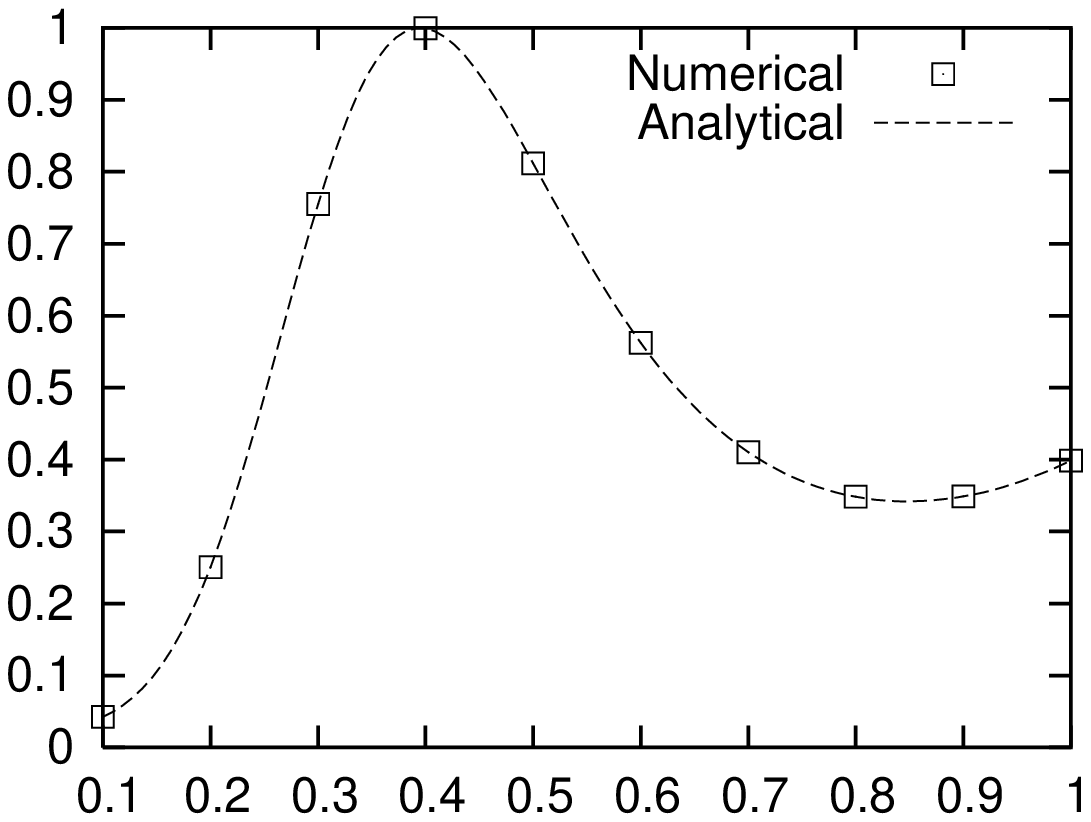}
\includegraphics[height=0.49\textwidth,keepaspectratio=,angle=270]{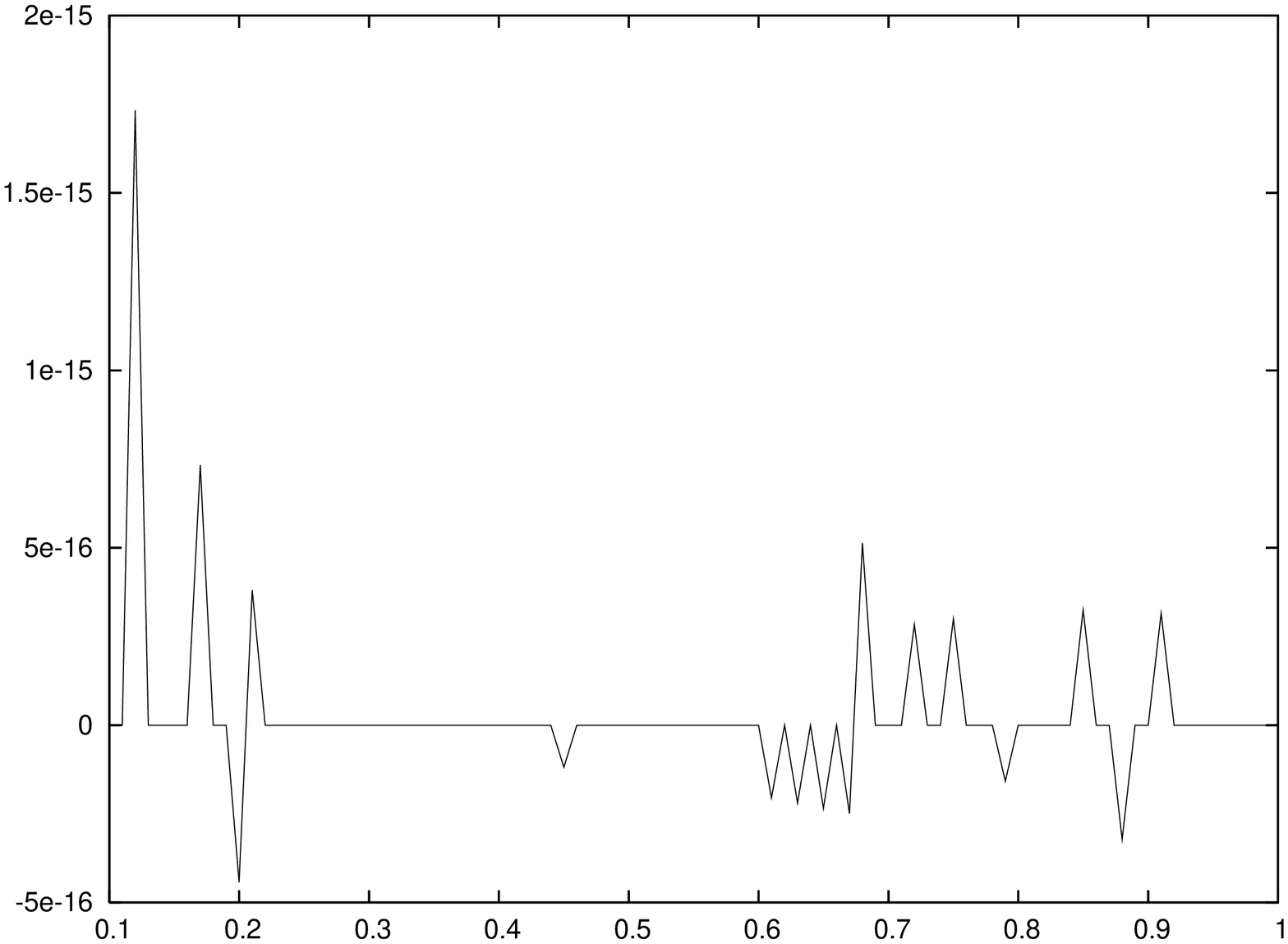}\\
\includegraphics[height=0.49\textwidth,keepaspectratio=,angle=270]{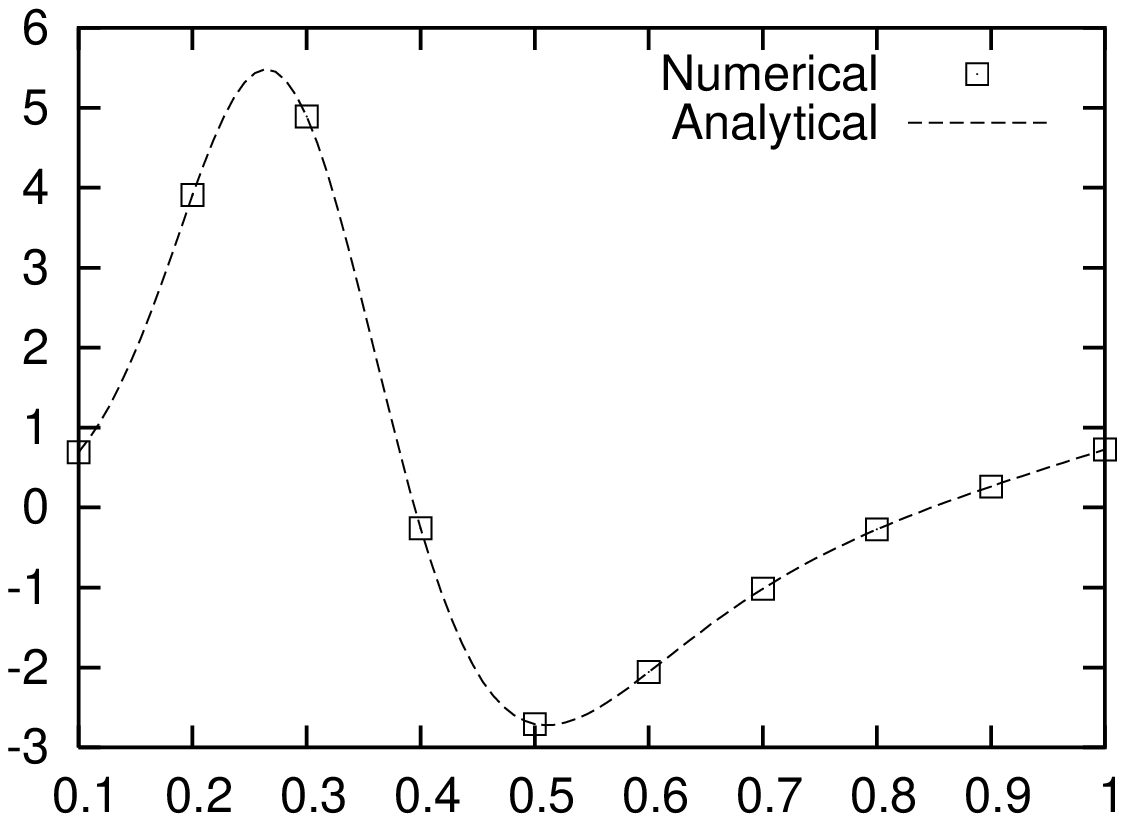}
\includegraphics[height=0.49\textwidth,keepaspectratio=,angle=270]{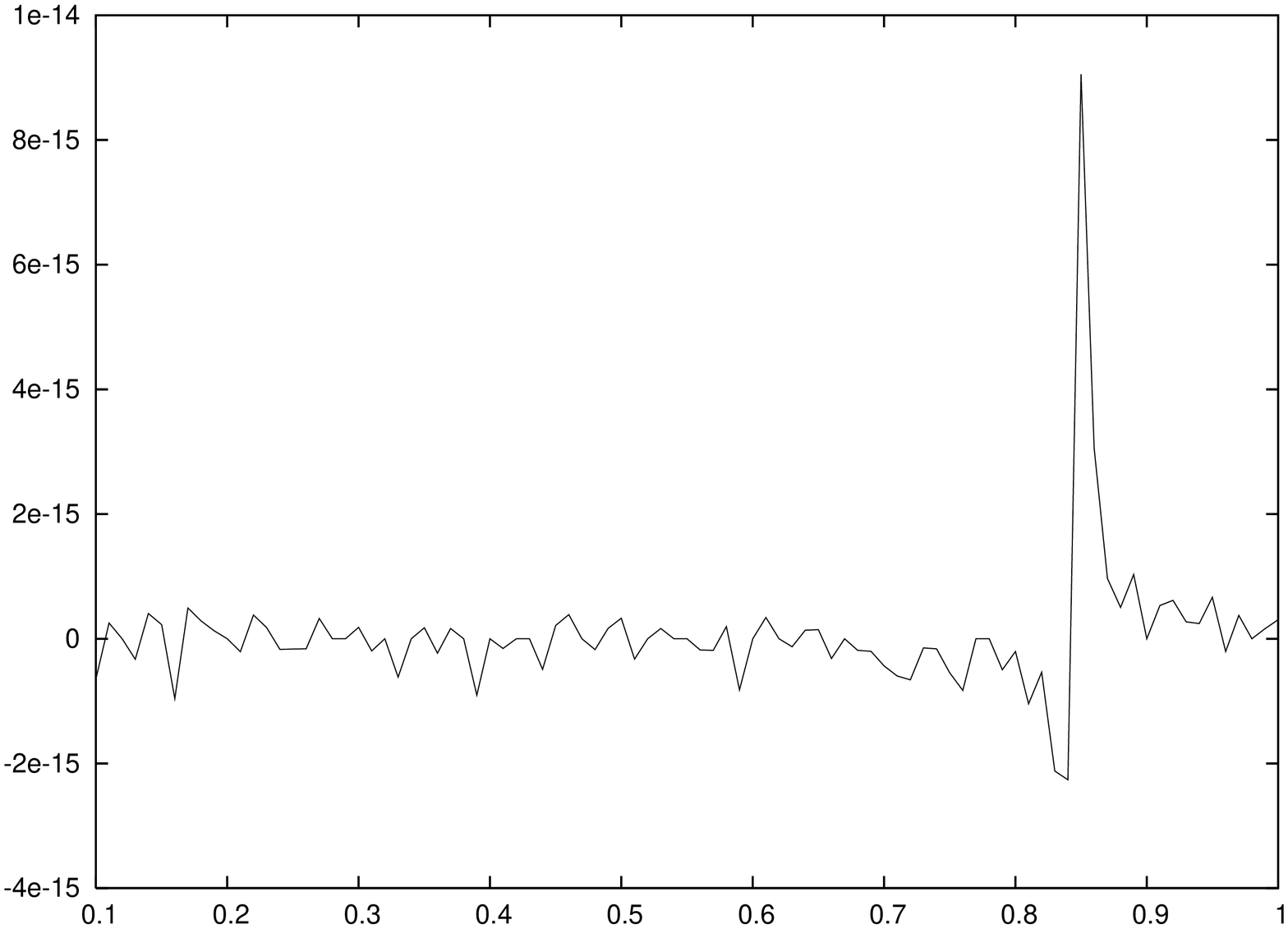}\\
\includegraphics[height=0.49\textwidth,keepaspectratio=,angle=270]{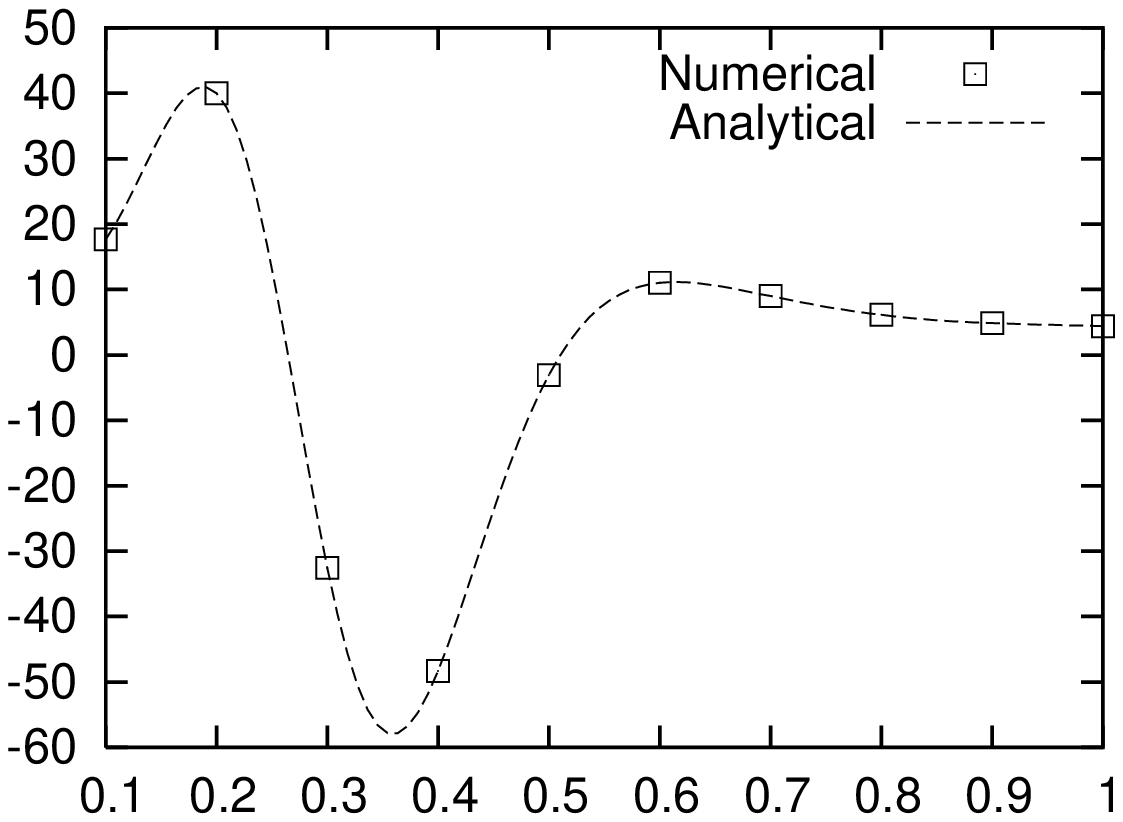}
\includegraphics[height=0.49\textwidth,keepaspectratio=,angle=270]{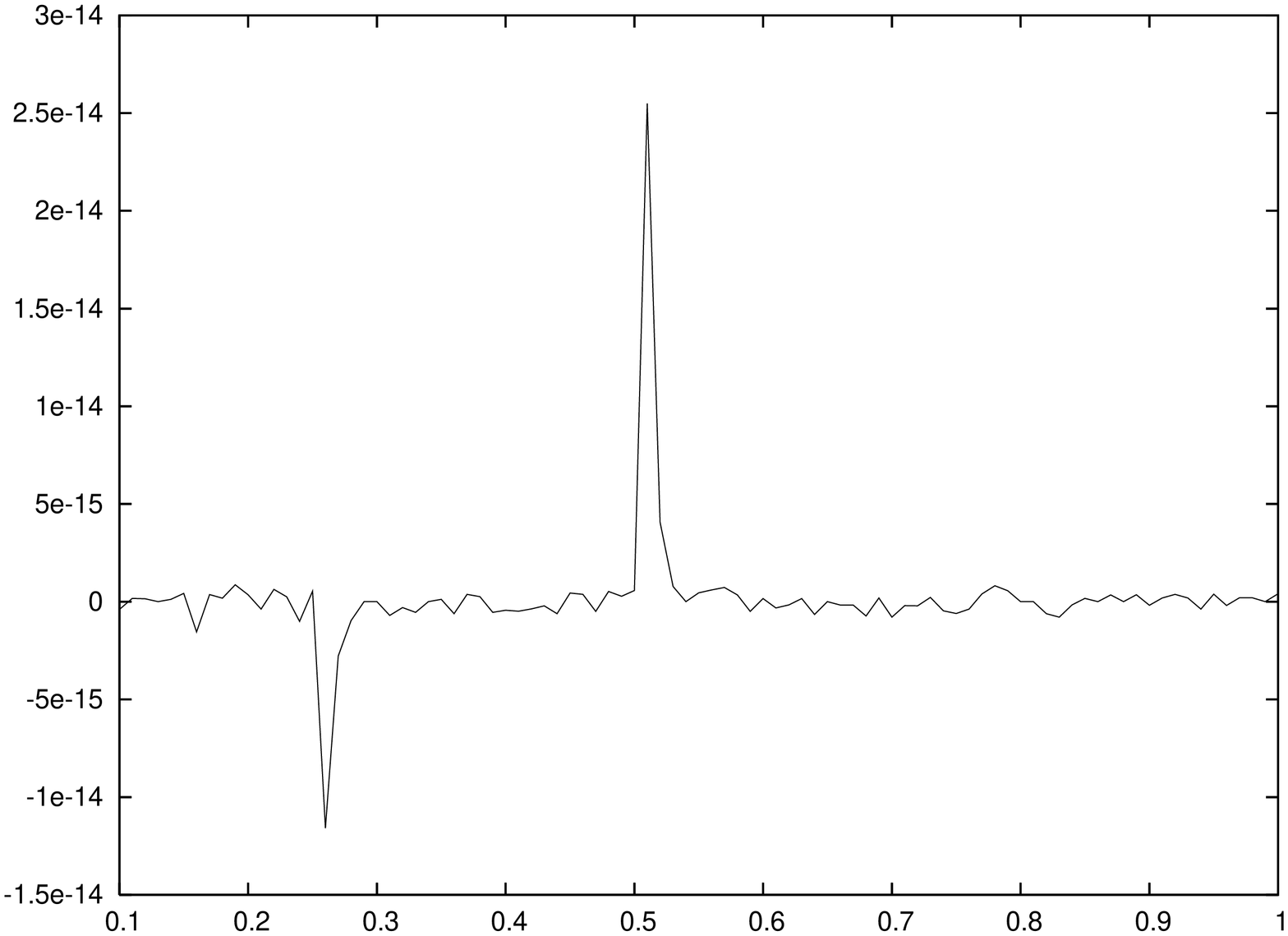}\\
\includegraphics[height=0.49\textwidth,keepaspectratio=,angle=270]{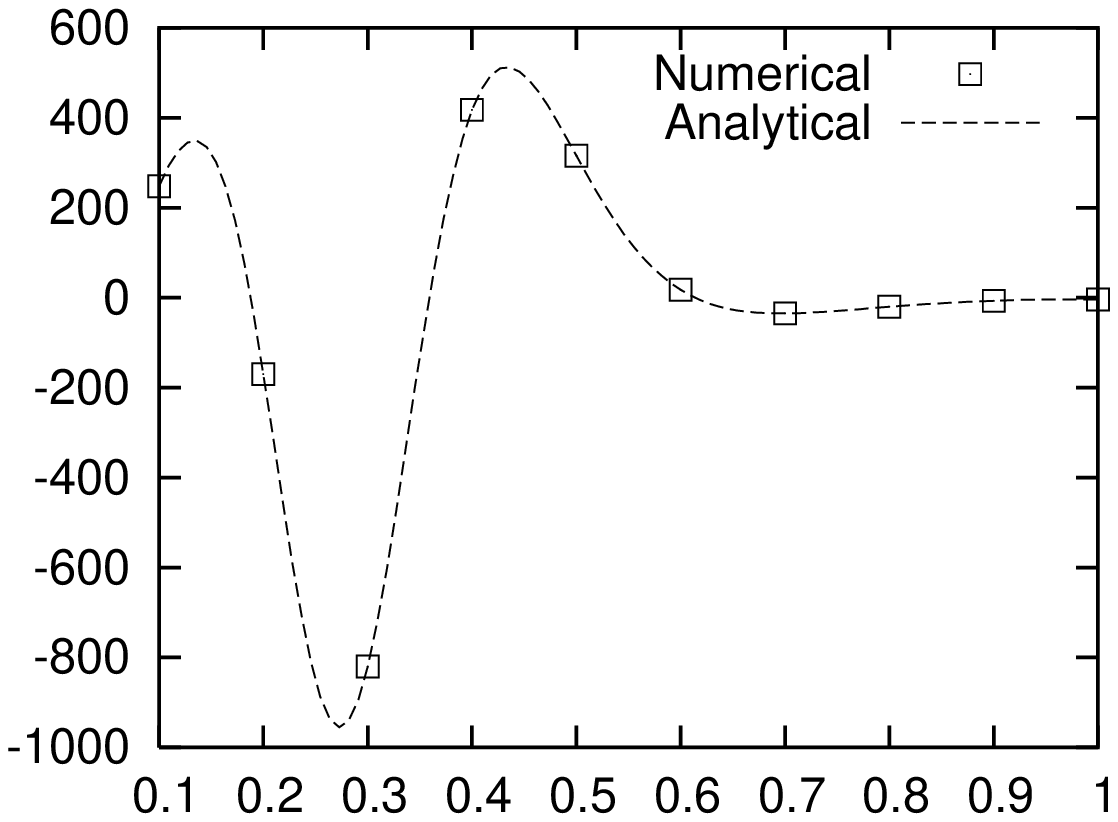}
\includegraphics[height=0.49\textwidth,keepaspectratio=,angle=270]{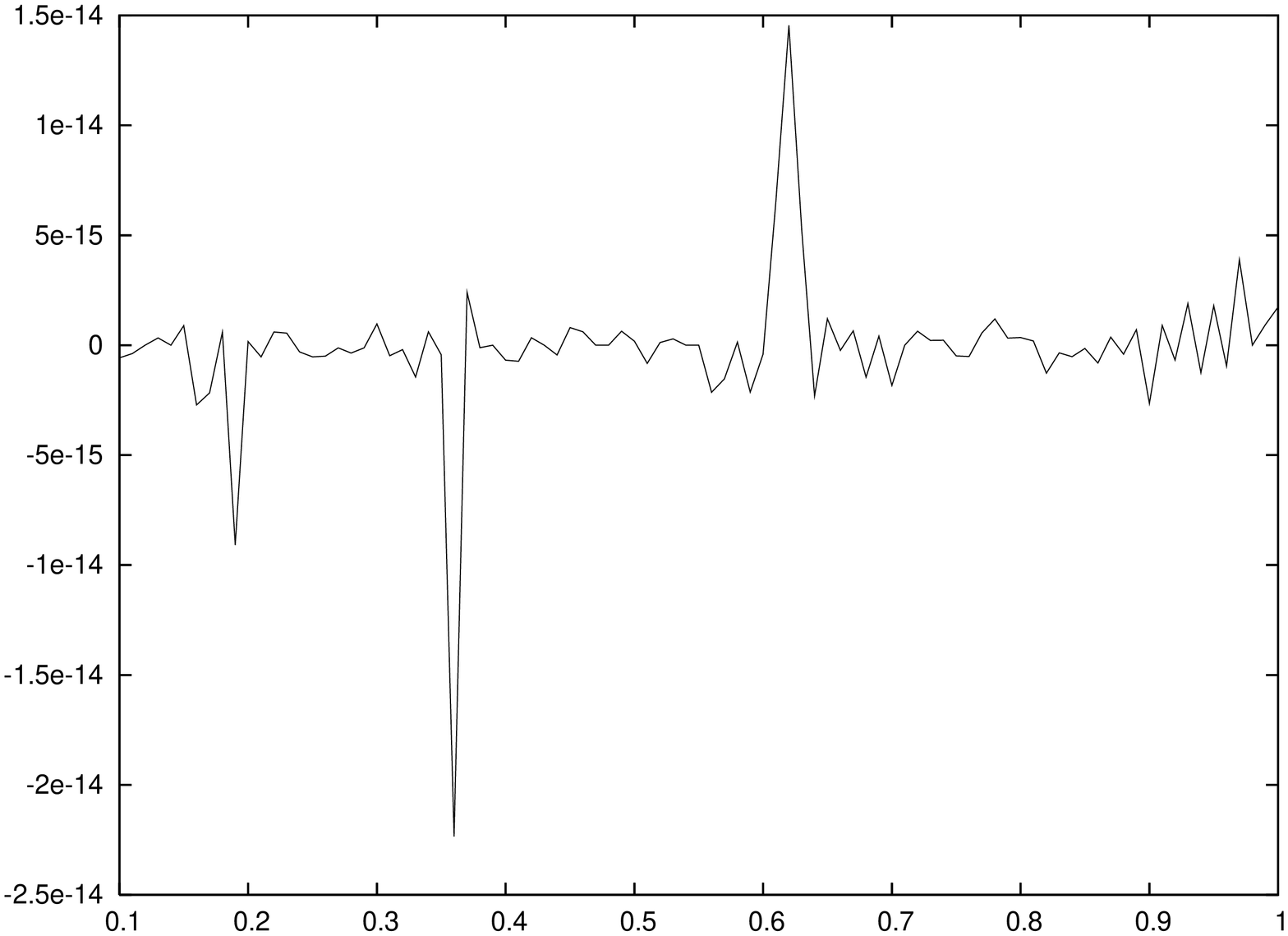}\\
\caption{Analytical (lines) and numerical (boxes) evaluations (left)
  and relative errors (right) for (from top to bottom) the
  function $\exp\Big(-\sin^2(5\sqrt{x})/\sqrt{x}\Big)$ and
  its first, second and third derivatives.}
\label{fig:one}
\end{figure}

As an example, let us consider the function
\begin{equation}
  f(x) = \exp\Big(-\sin^2(5\sqrt{x})/\sqrt{x}\Big)
\end{equation}
whose derivative expansion involves recourse to Fa\`a di Bruno's
formula twice at each order, as well as usage of the expansion of the
square root, power and multiplicative inverse
functions. Fig. \ref{fig:one} shows plots of the analytical form (as
given by a computer algebra program) and numerical evaluations using
\URAD\/ of $f(x)$ and it first three derivatives. Excellent agreement
can be seen, with the largest relative errors of order $10^{-14}$.

As a further test, a comparison was run between \URAD\/ and
ADOL-C. The example {\tt powexam} included with ADOL-C, which computes
the Taylor series coefficients for the monomial $x^n$, was used as a
test case. Complete agreement between \URAD\/ and ADOL-C was
found. For low orders of $n<10$, \URAD\/ was about 2-3 times faster
than ADOL-C, whereas for larger orders, ADOL-C began to take over
quickly. As a further comparison point, the {\tt speelpenning} example
program, which computes the function value and gradient of the product
$\Pi_{i=0}^{n-1}x_i$, was used. Again, agreement (to within floating
point accuracy) was found between \URAD\/ and ADOL-C. In this case,
\URAD\/ was about 5-10 times faster than ADOL-C for all dimensions in
the range 2$\dots$100 tested. These timings suggest that \URAD\/ is
very competitive for relatively low Taylor expansion orders (less than
about 10), but falls behind for higher orders. The \URAD\/
versions were the most direct possible translations of the ADOL-C
examples, and the timing tests were run on a $2.8$ GHz Intel Pentium 4
using the Intel Fortran compiler and GCC to compile the \URAD\/ and
ADOL-C versions, respectively.

\section{Summary}
\label{sect:summ}

\URAD\/ provides a much needed high-order automatic differentiation
package for \FOR\/ that is particularly useful for expansion of
Feynman diagrams in external momenta or particle masses, but is
expected to be applicable to other fields of computational physics as
well.

\section*{Acknowledgements}

The author acknowledges helpful discussions with A.G. Hart and
R.K. Lewis. This work was supported in part by the Natural Sciences
and Engineering  Research Council of Canada and by the Government of
Saskatchewan.


\section*{Test Run Output}

To verify that the \URAD\/ package has been installed and compiled
correctly, the user should build the included verification program by
performing a {\tt make verify} in the installation directory. When run,
{\tt ./verify} should produce the following output:
\begin{verbatim}
    Taylor eval.      Analytic eval.      Error
    0.333333333333    0.333333333333      0.000000000000E+00
   -0.888888888889   -0.888888888889      0.000000000000E+00
    1.185185185185    1.185185185185      0.000000000000E+00
   -2.370370370370   -2.370370370370      0.000000000000E+00
    6.320987654321    6.320987654321      0.000000000000E+00
  -21.069958847737  -21.069958847737      0.355271367880E-14
   84.279835390946   84.279835390947      0.142108547152E-13
\end{verbatim}
The less significant digits and errors may depend on the specific
floating-point implementation of the system used, but the two columns
headed {\tt  Taylor eval.} and {\tt Analytic eval.} should agree to
the accuracy shown (in double precision arithmetic).

If this verification passes, a number of additional tests can be
performed by performing {\tt make test} and running {\tt ./test},
which should print
\begin{verbatim}
 All tests passed!
\end{verbatim}
Any other output, in particular complaints about failed tests, is
indicative of a compilation error or a problem with the floating-point
system used, such as a lack of precision in the results of some of the
intrinsic functions.


\end{document}